\documentclass[twocolumn,showpacs,prb,superscriptaddress]{revtex4}
\newcommand{\figurewidth}{\columnwidth}

\usepackage{graphicx}
\def\bea{\begin{eqnarray}}
\def\eea{\end{eqnarray}}
\def\a{\alpha}
\def\d{\delta}
\def\p{\partial} 
\def\nn{\nonumber}
\def\r{\rho}

\def\la{\langle}
\def\ra{\rangle}
\def\e{\epsilon}
\def\o{\omega}

\def\g{\gamma}
\def\l{\lambda}

\def\f{\frac}
\def\dg{\dagger}
\def\zh{\hat{Z}}

\begin{document}

\title{Quantum transport using the Ford-Kac-Mazur formalism.}
\author{Abhishek Dhar $^{1,\dagger}$ and B. Sriram Shastry $^{2,\star}$}
\address{  $^1$ Physics Department, University of California, Santa Cruz,
  CA 95064 \\ 
$^2$ Indian Institute of Science, Bangalore 560012, India. }                   \date{\today}

\begin{abstract}
The Ford-Kac-Mazur formalism is used to study 
quantum transport in (1) electronic and (2) harmonic oscillator systems
connected to general  reservoirs. It is shown that for non-interacting
systems the method is easy to implement and is used to obtain many
exact results on electrical  and thermal transport in one-dimensional
disordered  wires. Some of these have earlier been obtained using
nonequilibrium Green function methods. We examine the role that
reservoirs and contacts can have on determining the transport
properties of a wire and find several interesting effects. 
\end{abstract}

\pacs{PACS numbers: 05.60.-k, 72.10.Bg, 73.63.Nm, 05.40.-a}
\maketitle


\section{Introduction}

There is considerable current interest in the problem of transport
through various nanoscale devices both from the fundamental and from
applied points of view. In this connection, Kubo's transport formulas
have to a large extent been superseded by  different formalisms in the
spirit of Bardeen's tunneling model \cite{bard}.
The  Landauer formula\cite{land}(LF) and the Keldysh
technique\cite{caroli}, quantum Langevin equations\cite{hu},  $C^*$ algebraic
formulas\cite{tasaki} and 
generalized scattering theory ideas\cite{todorov} have been developed,
  allowing one to study systems in steady state arbitrarily far from
the linear region where Kubo is applicable. 
There is also considerable experimental  activity  involving resistive
elements, such as quantum dots, STM tips, single walled nanotubes and
insulating nanowires, often coming up with unexpected physics \cite{white,kong,schwab}.  

The most popular alternative to Kubos formulas is the  LF,  proposed
in 1957 \cite{land}. Since then several derivations of the LF have
been given \cite{butt} and this has led to a  good understanding of
the formula. A large number of experiments are  interpreted
successfully on the basis of LF. The quantum of conductance $e^2/h$ has
been understood as a contact resistance which arises due to the
squeezing of the reservoir degrees of freedom into a single
channel \cite{shar,aaro}. While a physically careful statement of the
conditions for validity of the LF can be found in
Ref(\cite{landauerimryrmp}), we believe that  a detailed mathematical
theory  of the role  
of reservoirs and the nature of the coupling between the wires and
reservoirs does not exist.  The role of the idealized reservoirs has been to
serve as perfect sources and sinks of thermal electrons.
This clearly 
will not be satisfied in all experimental conditions and it is
necessary to have a better microscopic understanding of reservoirs and
contacts. 
There has been some work \cite{caroli,tasaki,todorov,aaro,meir} in
this direction but, to our knowledge, a detailed understanding of the
role of reservoirs is still lacking.

In this paper we adapt a formalism that was developed by Ford, Kac,
and Mazur \cite{ford} (FKM) and model reservoirs as infinite
non-interacting systems.  
This method was originally devised to study Brownian motion in coupled  
oscillators \cite{ford} and was later extended to a general study of
the problem of a quantum particle coupled to a quantum mechanical heat
bath \cite{lewis}.  
In this approach reservoirs are modelled by a collection of
oscillators which are initially in equilibrium. The reservoir degrees
of freedom are then eliminated leading to quantum Langevin equations
for the remaining degrees of freedom (the system). Thus the reservoirs
can be viewed as providing sources of noise and dissipation into the
system. The FKM formalism is thus very direct to interpret and, as we
shall demonstrate, is more straightforward to apply than other
methods of treating open quantum systems such as the
Caldeira-Leggett\cite{cald},  
Keldysh \cite{keld} and scattering theory\cite{todorov}.  
Quantum Langevin equations have earlier been used in the context of
transport in mesoscopic systems and have helped in the understanding of
some experimental data \cite{hu,clel}. The FKM approach was also used earlier
by O'Connor and Lebowitz \cite{conn}   in studying   classical
heat transport in   disordered harmonic  chains and our   analysis
here closely  follows theirs.  
Here we use the FKM approach to make a detailed study of quantum
transport in disordered electronic and phononic systems.    
For very general reservoirs we obtain exact formal expressions for
currents and local  densities in the nonequilibrium steady state.
We find that for a special type of
reservoir, the ideal Landauer result (where the conductance is expressed in
terms of the transmission coefficient of one-dimensional plane waves)
follows exactly, while for general reservoirs they need to be modified. 
We examine in some detail the effect on transport properties that the
choice of reservoirs can have and find a number of interesting effects.
For example in the electron case we find that imperfect contacts can
lead to an enhancement of conductivity. In the phonon case we find the
surprising result, earlier noted for classical systems, that the heat
current $J$ in a long disordered wire decays with system size $N$ as
$J \sim 1/N^{\alpha}$ where $\alpha$ depends on the low frequency spectral
properties of the reservoirs.  

The paper is organized as follows: In Sec.I we present the formalism
and results  for transport in the one dimensional Anderson model. In
Sec.II  we present the formalism and results for transport in disordered
harmonic chains. We end with a discussion in Sec.III.

\section{Transport in the one-dimensional Anderson model}
\subsection{The formalism and main results}
\begin{figure}
\includegraphics[width=\figurewidth]{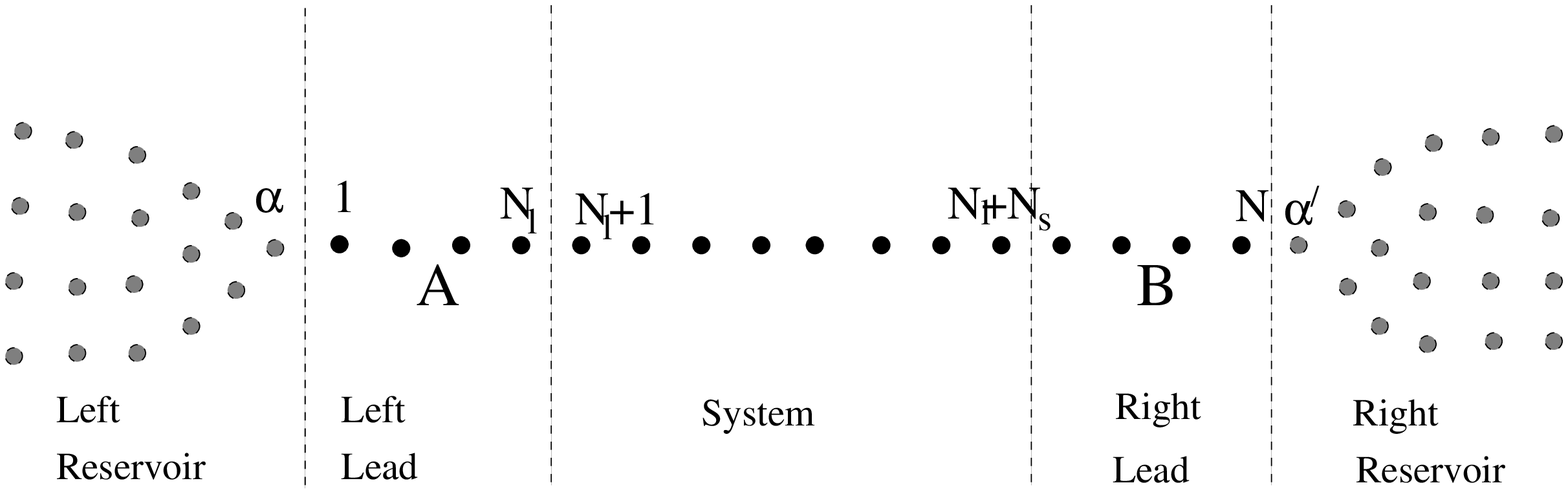}
\caption{~A disordered system connected through $1D$ leads
to reservoirs at different potentials and temperatures. 
\label{set}}
\end{figure}

The set-up:  We wish to study conduction in a disordered fermionic system
 connected to heat and particle reservoirs through ideal
 $1D$ leads [see Fig.~\ref{set}]. We consider a tight-binding model
 and for simplicity we take the system and leads to be  
 $1D$  while the reservoirs are quite 
general. 
We use the  following notation: the indices $l,m$ denote points on
 the system or leads, greek indices $\l,\nu$ or $\l', \mu'$ 
 denote points on the left or right reservoirs respectively, finally
 $p,q$ denote points anywhere. Thus $c_l~(l=1,2...N)$
 denotes lattice fermionic  operators on the (system + lead),
 $c_\l,~c_{\l'}~(\l,\l'=1,2...M)$  denotes 
 operators on the left and right reservoirs.  The $c_p$s' satisfy the usual 
anticommutation relations $\{c_p,c_q \}=0;~\{ c^{\dg}_p,c^{\dg}_q
\}=0;~\{ c^{\dg}_p,c_q \}=\d_{pq}$.  Out of the $N=N_s+2 N_l$, sites the first 
and
 last $N_l$ sites refer to the leads while
 the middle $N_s$ sites refer to the system.   
The Hamiltonian for the entire system  is given by $H =H^0+V +
 V_{int}$, where 
\bea
&&H^0 =  -\sum_{l=1}^{N-1}  (c^{\dg}_{l} c_{l+1} +c^{\dg}_{l+1} c_l )+
\sum_{l=1}^{N} v_l c^{\dg}_l c_l +\sum_{\l \nu} \hat{T}_{\l \nu}
 c^{\dg}_{\l} c_{\nu}+ \nn \\ 
&&\sum_{\l ' \nu'}\hat{T}'_{\l' \nu'} c^{\dg}_{\l'}
 c_{\nu'};~ 
V = -\g (c^{\dg}_1 c_{\a}+c^{\dg}_{\a} c_1) -\g' (c^{\dg}_N
 c_{\a'}+c^{\dg}_{\a'} c_N) \nn 
\eea
The first part of $H^0$ refers to the  system and
 leads, while $\hat{T}$ and $\hat{T}'$ describe the two reservoirs.
The  contact between the  reservoirs and leads is given by the
 interconnection  part $V$. The interaction part $V_{int}$
 can be added perturbatively, and we
 return to its inclusion  later in the paper.
We will consider a system with onsite disorder and so choose the onsite
 energies $v_l$,  $l=N_l+1...N_l+N_s$, from some random
 distribution. At  sites  belonging to the 
leads [$l=1,2,...N_l,N_l+N_s+1,...N$],  assumed to be perfect
 conductors, we set $v_l=0$.  
At some time $t < \tau$ in the remote past, the two
reservoirs are isolated and in equilibrium at chemical potentials $\mu$
and $\mu'$ and inverse temperatures $\beta$ and $\beta'$ respectively. 
At $t=\tau$, we connect the reservoirs to the two leads and
evolve the system with the Hamiltonian $H$.
We study the properties of the  nonequilibrium steady state, reached
 after a long time.  
 
The Heisenberg equations of motion for the operators
of the system and leads is given by (for $t > \tau$):
\bea
\dot{c}_1 &=& i c_2-i v_1 c_1 + i \gamma c_{\a} \nn \\
\dot{c}_l &=& i ( c_{l-1}+c_{l+1} ) -i v_l c_l~~ ( 2 \leq l \leq N-1)
\nn \\ 
\dot{c}_N &=& i  c_{N-1}-i v_N c_N +i \gamma' c_{\a'}.
\label{geqmot}
\eea
 The
equations at the boundary sites 
involve reservoir operators, $c_\a,~c_{\a'}$. Using the equations of
motion of the 
reservoir variables we can replace these reservoir operators 
by {{\it Langevin type}} terms. The equations for the left reservoir 
are given by (for $t > \tau$): 
\bea
\dot{c}_{\l} &=& -i  T_{\l \nu} c_{\nu} ~~ (\l \neq \a) \nn \\   
\dot{c}_{\a} &=& -i  T_{\a \nu} c_{\nu} +i \g  c_1.
\eea  
This is a linear set of equations with an inhomogeneous part given by
the term $i \g  c_1 $ and has the general solution 
\bea
c_\l(t) = i\sum_{\nu} g^+_{\l \nu}(t-\tau) c_\nu(\tau) - \int_\tau^\infty
dt' g^+_{\l \a}(t-t') (  \g c_1(t') ) \nn \\ ~~{\rm where }~~~~ 
g^+_{\l \nu}(t) =  -i \theta(t) \sum_n \psi_n(\l) \psi^*_n(\nu) e^{-i
  \e_n t}. \nn
\eea
Here $\psi_n(\l)$ is the single particle eigenstate of the left 
reservoir, with energy $\e_n$,  and $n$ runs over all states.
We need $c_{\a}(t)$ which we note has two parts. The first,
$h(t)=i \sum_{\nu} g^+_{\a \nu}(t-\tau) c_{\nu}(\tau)$, is like a
noise term whose statistics is determined by the initial
conditions of the reservoir. Initially the reservoirs are 
in thermal equilibrium and  
the normal modes $c_n=\sum_{\l} c_{\l} \psi_n(\l)$ satisfy $\la
c^{\dg}_n(\tau) c_{n'}(\tau) \ra  = \d_{n n'} f(\e_n,\mu, \beta) $, 
where $f$ is the Fermi distribution $f=1/[e^{\beta(\e_n-\mu )}+1]$
and $\la \hat{O} \ra=Tr[\hat{O} \hat{\rho}]$ where $\hat{\rho}$ is the
reservoir density matrix at time $\tau$ and $Tr$ denotes a trace over
reservoir variables.  
The second part of $c_{\a}(t)$, $-\g \int_{\tau}^{\infty} dt' g^+_{\a
\a}(t-t')  c_1(t')$, is dissipative in nature.  
Defining the Fourier transforms $c_{p }(\o)= \f{1}{2 \pi}
\int_{-\infty}^{\infty} d t c_{p}(t) e^{i \o t}$, $g^+_{\a \a}(\o)=
\int_{-\infty}^{\infty} d t g^+_{\a \a}(t) e^{i \o t}$ and $h(\o)= \f{1}{2 \pi}
\int_{-\infty}^{\infty} d t h(t) e^{i \o t}$,  
and taking limits $M \to \infty$ and $\tau \to -\infty$ we get:
\bea 
&& c_{\a}(\o)  
= h(\o) - \g g^{+}_{\a \a} (\o) c_1(\o)    \nn \\
&&\la {h^{\dagger}}(\o) h(\o') \ra = I(\o)\d (\o - \o'), \nn \\
&&  I (\o) =\rho_{\a}(\o) f(\o), \nn \\
&& g^{+}_{\a \a}(\o) = \sum_n \f{\mid \psi_n(\a) \mid^2}{\o-\e_n}- i
 \pi \rho_{\a}    
\label{spect} 
\eea
where  $\rho_{\a}=\sum_n \mid \psi_n(\a)  \mid ^2  \d (\o-\e_n)  $ is
the  density of states at site $\a$. The third equation above is a
statement of the fluctuation-dissipation theorem. 
Similarly for the right reservoir we get $ c_{\a'}(\o)  
= h'(\o) - \g' g^{+}_{\a' \a'} (\o) c_N(\o) $, with the noise
statistics of $h'(\o)$  determined by $\mu'$ and $\beta'$ . 
Also $h$ and $h'$ are independent so that $\la
h^{\dg}(\o) h'( \o') \ra = 0$.     
We now Fourier transform the system equations and plug in the
forms of $c_{\a}(\o)$ and $c_{\a'}(\o)$ to get the following particular
solution:
\bea  
c_l(t)&=& \int_{-\infty}^{\infty} d\o  \zh^{-1}_{lm}(\o) h_m(\o) e^{-i
\o t}  \label{sol} \\
\zh_{lm} &=& \hat{\Phi}_{lm}+\hat{A}_{lm} \nn \\
\hat{\Phi}_{lm} & =&  -\d_{l,m+1}-\d_{l,m-1}+(v_l-\o) \d_{l,m} \nn \\ 
\hat{A}_{lm} &=& \d_{l,m}[\g^2 g^+_{\a \a} (\o)  \d_{l,1}+ {\g'}^2
g^+_{\a' \a'}(\o) \d_{l,N} ] \nn \\
h_l &=& \g h(\o) \d_{l,1}+ \g' h' (\o) \d_{l,N}. \nn
\eea
 With this formal solution and the known properties of the
spectral functions $h(\o)$, $h'(\o)$, $g^+_{\a \a}(\o)$ and $g^+_{\a'
\a'}(\o)$, we can now  
compute various physical quantities of
interest. Specifically we shall be interested in the electrical and
thermal currents and the local particle and energy densities. 
The operators corresponding to particle and energy densities are
given by 
\bea
\hat{n}_l &=& c^{\dg}_l c_l \nn \\
\hat{u}_l &=& -(c^{\dg}_{l} c_{l+1} +c^{\dg}_{l+1} c_l )+
\f{1}{2} ( v_l c^{\dg}_l c_l +v_{l+1} c^{\dg}_{l+1} c_{l+1}) 
\eea
while the corresponding current operators $\hat{j}^n$ and $\hat{j}^u$
are defined through the conservation equations $\p \hat{n}/\p t +\p
\hat{j}^n/\p x =0$ and $\p \hat{u}/\p t +\p \hat{j}^u/\p x =0$. We
get:
\bea
\hat{j_l}^n &=& i(c^\dg_{l+1} c_l- c^\dg_l c_{l+1}) \nn \\
\hat{j_l}^u &=& -i  (c^\dg_{l+2} c_l- c^\dg_l c_{l+2})+
\f{v_{l+1}}{2}(\hat{j}^n_{l+1}+  \hat{j}^n_{l})  
\label{curdef}
\eea

We now calculate the steady state averages of these four quantities.
We introduce some notation and state a few
mathematical identities.
We denote by $Y_{l,m}$ the determinant of the
submatrix of $\zh$ beginning with the $l$th row and column and ending
with the $m$th row and column. Similarly $D_{l,m}$ denotes determinant
of the submatrix formed from $\hat{\Phi}$.
The following results can be proved: $
(i)~Y_{1,N}= D_{1,N}+\g^2 g^{+}_{\a \a} D_{2,N}+{\g'}^2 g^{+}_{\a' \a'} 
D_{1,N-1}
+\g^2 {\g'}^2 g^{+}_{\a \a} g^{+}_{\a' \a'} D_{2,N-1};~
(ii)~\zh^{-1}_{lN}=Y_{1,l-1}/Y_{1,N};~\zh^{-1}_{l1}=Y_{l+1,N}/Y_{1,N};
(iii)~D_{1,n-1} D_{2,n}-D_{1,n}D_{2,n-1}=1. $

{\it{Particle and Heat Currents:}}
The expectation value of the current
operators, using Eqns.~(\ref{spect},\ref{sol}), gives: 
\bea
&&\la j_l^n \ra= -2 \int_{-\infty}^{\infty} d\o Im[ \sum_{r=1,N}
\zh^{-1*}_{l+1, r}(\o) 
\zh^{-1}_{l, r}(\o) I_r(\o)] \nn  \\
&& \la \hat{j}^u_l \ra=2 \int_{-\infty}^{\infty} d\o Im[ \sum_{r=1,N} \zh^{-1*}_{l+2, 
r}(\o)
\zh^{-1}_{l, r}(\o) I_r(\o)], 
\label{cureq}
\eea
where $I_1=\g^2 I;~ I_N={\g'}^2 I'$. 
In the case of the heat current we take 
$l$ to be on the leads so that $v_l=0$.
Using the various identities stated earlier we can show, as expected,
that these are independent of $l$ and reduce to the simpler
expressions 
\bea
\la \hat{j}^n_l \ra &=& \int_{-\infty}^{\infty}
d\o J(\o) [f(\o)-f'(\o)]  \nn \\
\la \hat{j}^u_l \ra &=&  \int_{-\infty}^{\infty}
d\o  \o J(\o) [f(\o)-f'(\o)]~~{\rm where} \nn \\
J(\o) &=&  2 \pi \g^2 {\g'}^2  \rho_{\a}(\o) \rho_{\a'}(\o)
/{\mid Y_{1,N} \mid^2}. \nn 
\eea  
These can be expressed in terms of the retarded Green's function
$G^{+}(\o)=(\o+i \e -H)^{-1}$. This 
satisfies $ G^+=g^++g^+ V G^+ $ where $g^+=(\o+i \e
-H^0)^{-1}$. These can be solved to give :
\bea
G^+_{1m} &=& [g^+_{1m}-{\g'}^2  g^+_{\a' \a'}( g^+_{NN}
g^+_{1m} -g^+_{1N} g^+_{Nm})]/Z \nn \\  
G^+_{Nm} &=&[ g^+_{Nm}-\g^2 g^+_{\a \a} ( g^+_{11} g^+_{N m}- g^+_{N1}
g^+_{1m} ) ]/Z~~{\rm where}~~ \nn \\ 
Z &=&1-\g^2 g^+_{11} g^+_{\a \a}-{\g'}^2 g^+_{NN} g^+_{\a'
\a'} \nn \\ && + \g^2 {\g'}^2 g^+_{\a \a} g^+_{\a' \a'} (
g^+_{11}g^+_{NN}-g^+_{1N} g^+_{N1} ) \nn 
\eea

Let $g_{lm}=Re[g^+_{lm}]$ denote the real part of the system's
Green function. It is easy to see that 
$g_{1N}=g_{N1}=-1/D_{1,N};~g_{11}=-D_{2,N}/D_{1,N};~g_{NN}=-D_{1,N-1}/D_{1,N}$.
Using these and the Jacobi identity $g_{11}g_{NN}-g_{1N}
g_{N1}=D_{2,N-1}/D_{1,N}$ we get  $1/{\mid Y_{1,N}
\mid^2}=Gn^+_{1N} Gn^-_{N1}$ where $Gn^+$ is a modified Green function
obtained from $G^+$ by replacing all {\it system} green functions by their
real part. We then get the particle current in a
form similar to those obtained by Meir and Wingreen \cite{meir} using
the Keldysh formalism and by Todorov et al \cite{todorov} using
time-independent scattering theory. Their results differ from
ours in that they are expressed in terms of $G^+$ 
instead of $Gn^+$. The case of insulating wires treated by Caroli et al 
\cite{caroli} also follows from our results.

{\it Scattering states:} 
It is instructive to write the  
currents and densities in terms of properties of the
single-particle scattering states of the full Hamiltonian $H$ ( possible when interactions are absent). 
Let $\psi^{jL}(\o)$ and $\psi^{jR}(\o)$ denote the $j$th unperturbed
wave functions  with energy $\o$, of the left and right
reservoirs respectively. 
Let  $a^{jL}_p$ and $a^{jR}_p$ denote the amplitude at site $p$ of the
$j$th right and left moving states obtained by evolving the
unperturbed levels with the full Hamiltonian. 
We then get: $a^{jL}_l=K^{-1}_{l1}\g
\psi^{jL}_{\a}(\o);~  
a^{jR}_l=K^{-1}_{lN}\g' \psi^{jR}_{\a'}(\o) $.
The currents and densities are given by $j^n_l=i(a^*_{l+1} a_l-a^*_l
a_{l+1})$, $n_l=a^*_l a_l$ etc.  Using these we find that $J(\o)$ 
is simply the total transmitted current for all waves
with energy $\o$. Also the particle density is given by
\bea
\la \hat{n}_l \ra= 
 \int_{-\infty}^{\infty} d\o [\r^L_l(\o) f(\o)+ \r^R_l(\o) f'(\o)], \nn
\eea
where $\r^L_l=\sum_j |a^{jL}_l|^2$ is the total particle density at a
point $l$ due to all right moving waves  with energy $\o$ and
$\r^R_l$ is 
due to left movers. 
Note that the currents and densities 
do not have the simple Landauer form since $J(\o)$ 
depends not only on the system but also on bath and contact
properties. The spectral properties of the baths enter into the
expressions in a nontrivial way and one cannot separate
the contributions of the system and the baths.

\subsection{ Ideal reservoirs and contacts:The Landauer case}
 This corresponds to the case where $\g=\g'=1$
and the reservoirs themselves are semi-infinite extensions of the
one dimensional leads. This results in reflectionless contacts between
the reservoirs and leads. The reservoir wave functions and energy
eigenvalues are: $\psi_n(\l)=[2/(M+1)]^{1/2} sin(k \l)$, $\e_n=-2
\cos(k)$ where $k=n \pi/(M+1)$ with $n=1,2,...M$. The leads are
connected at the end of the reservoir chains so that $\a=\a'=1$.   
We then get the following reservoir spectral functions:
\bea
&& I(\o) = \f{1}{\pi} [1-w^2/4]^{1/2} f(\o,\mu,\beta)~~|\o|<2;
\nn \\
&&I(\o)=0~~|\o|>2 \nn \\
&& g^+_{\a \a}(\o) = -e^{ik}, ~~-2 \leq \o=-2 \cos(k) \leq  2  \nn \\
&&~~~~~~ = \o/2  - sgn(\o) (\o^2/4-1)^{1/2},~~ |\o| > 2 
\label{bath}
\eea
We have similar expressions for the right reservoir. 
Let us use the notation that if sites $N_l+l$ and $N_l+m$ belong to the
system then we write $Y_{N_l+l,N_l+m}=y_{l,m}$, $D_{N_l+l,N_l+m}=d_{l,m}$. 
It can be shown  that the transmission probability of a wave
with momentum $k$ across the system is given by 
\bea
&& T=\f{4 \sin^2(k)}{|y_{1,N_s}|^2}~~ {\rm where} \label {trns}
\\
&& |y_{1,N_s}|=|d_{1,N_s}-e^{ik}(d_{2,N_s}+d_{1,N_s-1})+e^{i 2
k}d_{2,N_s-1}| 
\eea
Note that in this case the transmission factor {\it does not} involve
properties of the reservoirs and contacts. Also transmission is only
by propagating modes which can be labelled by a real wave vector $k$
(In general, non-propagating modes would also carry current and we
would have integrate over all frequencies).     
We then get the following forms for the particle and energy currents:
\bea
&& \la \hat{j}^n_l \ra= \f{1}{2 \pi} \int_0^{\pi} dk \nu(k) T(k)
[f-f']  \nn \\
&& \la \hat{j}^u_l \ra=\f{1}{2 \pi} \int_0^{\pi} dk \nu(k) \e(k)  T(k)
[f-f'],  ~~{\rm where} \label{cur} \\
&& \nu(k)=\p \e(k)/\p k=2 \sin(k).     \nn 
\eea
which are precisely of the  Landauer form.

In order to get the four-probe result we need to find the actual
potential and temperature differences across the
system. We imagine doing this by putting potentiometers and
thermometers at points on the leads [$A$ and $B$ in
Fig.~(\ref{set})]. These measure the local particle and 
energy density on the leads from which one can compute the chemical
potential and temperature.  We note that we do not expect local
thermal equilibration in this noninteracting system and so these are
only {\it effective} potentials and temperatures.

We start with the general expressions for densities [similar to
Eqns.~(\ref{cureq})] and after using the 
various determinantal identities we get ( for points $l$ located on
the left lead) an integrand which contains a factor
$\sin^2[k(N_l-l)]$. Assuming that $N_l$ is large and $l$ is not too close
to the point of contact with reservoirs  this factor can be replaced
by $1/2$. We then get for the particle and energy densities:
\bea
&& \la \hat{n}_l \ra = \f{1}{2 \pi} \int_0^{\pi} dk  \{ [2-T(k)]f+T(k)
f' \} \nn \\
&& \la \hat{u}_l \ra = \f{1}{2 \pi} \int_0^{\pi} dk  \e(k) \{
[2-T(k)]f+T(k) f' \}.
\label{dns}
\eea
We get similar expressions for densities at points on the right
lead. The expressions in Eqns.~(\ref{cur},\ref{dns}) are identical
to those obtained from semiclassical arguments, are true for ideal
contacts, and lead to the usual four-probe formulas.  The results of
Eqns.~(\ref{cur},\ref{dns}) have been obtained earlier by Tasaki
\cite{tasaki} using the theory of $C^*-$algebra.  
They can be easily extended to
the case where the leads are still one-dimensional but the system is of more 
general form. Thus let the system consist of $N_s$ points of which $1$ and
$N_s$ are connected to the two leads. Let us specify the system by the
matrix $\hat{\phi}$ such that $\hat{\phi}_{ll}=v_l-\o $ and
$\hat{\phi}_{lm}=-1$ whenever two distinct points $l$ and $m$ are
connected by a hopping element.
Then all the above formulas Eqns.~(\ref{cur},\ref{dns}) for
currents and densities hold provided we evaluate them within the leads
and  use the appropriate expression for the transmission
coefficient, namely
\bea
T= \f{4 \sin(k)^2 F^2}{|d_{1,N_s}-e^{ik}(d_{2,N_s}+d_{1,N_s-1})+e^{i 2
k}d_{2,N_s-1}|^2} 
\eea 
where $F$ denotes the determinant of the submatrix formed from $\hat{\phi}$
by deleting the $1$st row and $N_s$th column while $d$ is, as before,
but now constructed from $\hat{\phi}$.

\subsection{An application}
As an application we  show how
the experimental results of Kong et al \cite{kong} can be understood
qualitatively using our results by assuming imperfect contacts. 

\begin{figure}
\includegraphics[width=\figurewidth]{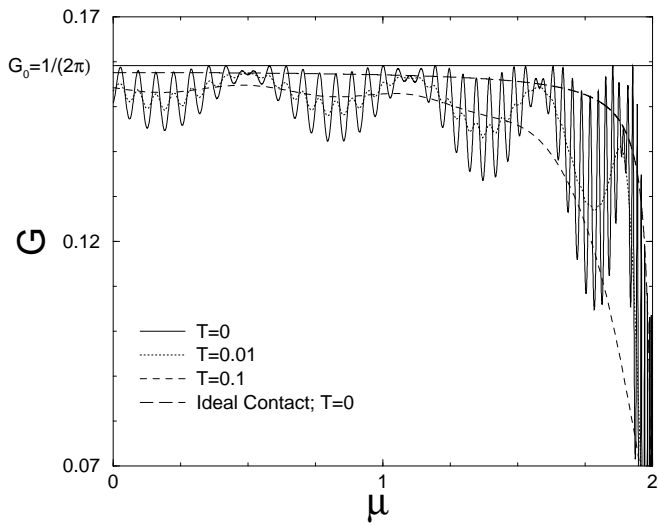}
\caption{ Plot of conductance-versus-Fermi level, at three different
temperatures (T), of a wire with a single impurity and imperfect
contacts. The corresponding plot for  perfect contacts
at $T=0$ is also shown. The horizontal line is the ideal
conductance $G_0=1/(2 \pi)$ [in units of $e^2/\hbar$]. 
\label{osc}}
\end{figure}

We consider again semi-infinite ideal reservoirs but make the contacts
non-ideal by setting $\gamma=\gamma'=0.9$. As system we take a wire
with a single impurity at site $s$ (Thus $v_s \neq 0$). The linear response
conductance is then given by $G= \int_{-\infty}^{\infty}
d\o  J(\o) f(\o)[1-f(\o)]$. 
We evaluate this numerically at different temperatures for $N=100$,
$s=10$ and $v_s=0.2$ [Fig.~(\ref{osc})]. We see the following
features: (a) a rapid oscillation of the conductance due to resonances
with standing waves in the wire, (b) a slower oscillation due to
standing waves formed between boundary and impurity and (c) washing
away of the oscillations with increasing temperature.   
These features are qualitatively the same as seen in the experiments
in \cite{kong}. The overall decrease in conductance with increasing
temperature is presumably due to scattering by phonons and hence is
not seen here. We have also plotted in Fig.~(\ref{osc}) the
conductance as given by the usual LF. 
Note that this {\it does not give the oscillatory
features}. Thus imperfect contacts cannot be treated as
resistances in series with the system. 
Another rather 
remarkable effect we see is the enhancement of the conductance as a result of
introduction of imperfect contacts. Infact we can see in
Fig.~(\ref{osc}) that at certain values of $\mu$ the
conductance almost attains the ideal value $1/(2 \pi)$. 
Similar features are also obtained if we make the contacts ideal but
take other forms of reservoirs (e.g rings or two
dimensional baths). 

\subsection{ Interacting systems}
 For this case the present approach readily
yields to a perturbative treatment. For illustration consider the case
where the Hamiltonian of the system (and lead) $H_{SL}$ contains an
interacting part and is given by
\bea
H_{SL}=-\sum_{l=1}^{N-1}  (c^{\dg}_{l} c_{l+1} +c^{\dg}_{l+1} c_l )+
\sum_{l=1}^{N} v_l c^{\dg}_l c_l+ \Delta \sum_{l=1}^{N-1} n_l n_{l+1}, \nn
\eea 
while the reservoirs are still taken to be non-interacting. In this
case, Eqns.~\ref{geqmot} take the form
\bea
\dot{c}_1 &=& i c_2-i v_1 c_1 -i \Delta n_2 c_1 + i \gamma c_{\a}  \nn \\
\dot{c}_l &=& i  ( c_{l-1}+c_{l+1} ) -i v_l c_l-i \Delta
(n_{l-1}+n_{l+1}) c_l, ~~~\nn \\ &&~~~~~~~~~~~~~~~~~~~2 \leq l \leq N-1 \nn \\
\dot{c}_N &=& i  c_{N-1}-i v_N c_N -i \Delta n_{N-1} c_N +i \gamma' c_{\a'}  
\label{geqmotI}
\eea
and, being nonlinear, can no longer be solved exactly. However it is
straightforward to obtain a perturbative solution which,
schematically, has the  form $c(\o)= \zh^{-1} h -\Delta \zh^{-1}
\int d\o'\int d\o'' \zh^{-1}h \zh^{-1}h \zh^{-1}h+O(\Delta^2)$. 
The operators for particle density and particle current remain
unchanged and we can obtain their expectation values as a
perturbation series using this solution.
Another possibility would be to solve Eq.~(\ref{geqmotI}) using a
self-consistent mean field theory.

\section{Heat Transport in oscillator chains}
We now use the FKM method to study heat conduction in quantum
disordered harmonic chains connected to 
general heat reservoirs which are modeled as infinite collection of
oscillators. There has been some earlier work on quantum wires
\cite{zurcher,saito} which follow a
similar approach but we give a more clear and complete picture and
make some interesting predictions for experiments.     

As in the electronic case we obtain formal exact
expressions for the thermal current and show that, for a special
case, they reduce to Landauer-like forms. We also analyse the 
asymptotic system size dependence of the current 
and show that, depending on the reservoirs, a long wire can
behave either like an insulator or a superconductor. 
Our results should be useful in  interpreting recent experiments
\cite{schwab} on heat transport in insulating nanowires and
nanotubes. They are also of interest in the context of the question of
validity of Fourier's law in 
one-dimensional systems, a problem that has received much attention
recently \cite{four}. A large amount of work on classical Hamiltonian
systems seem  to indicate that Fourier's law is not
valid in one-dimensionsal momentum conserving systems. 
Our work here shows that this is true even in quantum-mechanical
systems. 

\subsection{Formalism and main results}
We consider a mass disordered harmonic chain containing $N$ particles
with the following Hamiltonian:
\bea
H=\sum_{l=1}^N \frac{p_l^2}{2 m_l} +\sum_{l=1}^{N-1}
\frac{(x_l-x_{l+1})^2}{2}  +\f{( x_1^2+ x_N^2)}{2} \label{sysH}
\eea
where $\{x_l\}$ and $\{p_l\}$ are the displacement and momentum
operators of the particles  and  $\{m_l\}$ are the random masses.
Sites $1$ and $N$ are connected to two heat reservoirs ($L$ and $R$) which we now
specify. We model each reservoir by a collection of $M$
oscillators. Thus the left reservoir has the following Hamiltonian:
\bea
H_{L} &=& \sum_{l=1}^M \frac{P_l^2}{2 } + \sum_{l , m} \f{1}{2}K_{lm} X_l 
X_m  \label{resH} \\
&=& \sum_{s=1}^M \frac{\tilde{P}_s^2}{2}+ \f{\o_s^2}{2} \tilde{X}_s^2
\nn = \sum_{s=1}^M (n_s+1/2) \o_s a_s^\dagger a_s \nn,
\eea
where $K_{lm}$ is a general symmetric matrix for the spring couplings,
$\{X_l,P_l\}$ are the bath operators  and 
$\{\tilde{X}_l,\tilde{P}_l\}$  are the corresponding normal mode
operators. 
They are related by the transformation $X_l= \sum_s U_{ls}
\tilde{X}_s$ where  $U_{ls}$, chosen to be real, satisfies the
eigenvalue equation $\sum_l 
K_{nl} U_{ls} =\o_s^2 U_{ns}$ for $s=1,2...M$. The annihilation and
creation operators $a_s$, $a^{\dagger}_s$ are given by
$a_s=(\tilde{P}_s-i \o_s \tilde{X}_s)/(2 \o_s )^{1/2}$, etc. and
$n_s=a^\dagger_s a_s$ is the number operator.

The two reservoirs are initially in thermal equilibrium at
temperatures $T_L$ and $T_R$. At time $t=\tau$ the system, which is
in an arbitrary initial state is connected to the reservoirs. We
consider the case where site $1$ on the system is connected to $X_p$
on the left  reservoir while $N$ is connected to $X_{p'}$ on the right
reservoir. Thereafter the whole system evolves through the combined
Hamiltonian: 
\bea
H_T=H+H_L+H_R-k x_1 X_p - k' x_N X_{p'}. \nn
\eea

The Heisenberg equations of motion of the system variables 
are the following (for $t > \tau$):
\bea
m_1 \ddot{x}_1 &=& -[2 x_1-x_2] + k X_p \nn \\
m_l \ddot{x}_l &=& -(-x_{l-1}+2 x_l-x_{l+1}) ~~~1<l<N \nn \\
m_N \ddot{x}_N &=& -[-x_{N-1}+2 x_N]+k' X_{p'}.  
\label{seqm}
\eea
We note that they involve the bath variables $X_{p,p'}$. However
these can be eliminated and replaced by effective noise and
dissipative terms,  by using the equations of motion of the bath
variables. 
Consider  the equation of motion of the left bath variables. They
have the form:
\bea
\ddot{X}_n=-K_{nl} X_l ~~n \neq p \nn \\
\ddot{X}_p=-K_{pl} X_l +k x_1 \label{reseqm}
\eea
This is a linear inhomogeneous set of equations with the solution
\bea
X_n &=& \sum_l [F_{nl}(t-\tau) X_l(\tau)+ G_{nl}(t-\tau)
\dot{X}_l(\tau) \nn \\
&& +\int_\tau^{\infty} dt' G_{np}(t-t') k x_1(t') ~~{\rm where} \label{bsol} \\
F_{nl}(t)&=&\theta(t) \sum_s U_{ns} U_{ls} \cos(\o_s t);\nn \\
~G_{nl}(t) &=& \theta(t) \sum_s U_{ns}
U_{ls} \f{\sin(\o_s t)}{\o_s}. \nn 
\eea
Thus we find that $X_p$ (say) appearing in Eq.~(\ref{seqm})
has the form $X_p(t)= h(t) + k \int_{\tau}^{\infty} dt'
G_{pp}(t-t') x_1(t') $. The first part, given by  $h(t)=\sum_l [F_{pl}(t-\tau)
X_l(\tau)+ G_{pl}(t-\tau)\dot{X}_l(\tau) $, is like a noise term while
the second part  is like
dissipation. The noise statistics is easily obtained using the fact
that at time $t=\tau$ the bath is in thermal equilibrium and
the normal modes satisfy $\la a^\dagger _s(\tau) a_{s'}(\tau) \ra =f(\o_s,\beta_L) \d_{ss'}$.
 Here $f=1/(e^{\beta \o}-1)$ is the equilibrium phonon
distribution and $\la \hat{O}\ra=Tr[\hat{\r} \hat{O}] $ where
$\hat{\r}$ is the reservoir density matrix and $Tr$ is over the
reservoir degrees of freedom. We define the Fourier transforms:
$x_l(\o)=\f{1}{2\pi} \int_{-\infty}^{\infty} dt x_l(t) e^{i \o t}$,
$G^{+}_{pp}(\o)= \int_{-\infty}^{\infty} dt G_{pp}(t) e^{i \o t}$ and $h(\o)=
\f{1}{2 \pi} \int_{-\infty}^{\infty} dt h(t) e^{i \o t}$. Taking limits $M \to \infty$ and $
\tau \to -\infty$ we get:
\bea
&& X_p(\o)=h(\o)+k G^+_{pp}(\o) x_1(\o) \label{noise} \nn \\
&&\la h(\o) h(\o') \ra = I(\o) \d (\o+\o') \nn  \\
&&I(\o)=\f{f(\o) b(\o)}{\pi} \nn \\
&&G^+_{pp}(\o)=\sum_s \f{U^2_{ps}}{\o^2_s-\o^2}-i b(\o) ~~~{\rm where}
 \\
&&b(\o)=\sum_s \f{\pi U^2_{ps}}{2 \o_s} [\d (\o-\o_s) -\d (\o+\o_s)] \nn 
\eea 
Similarly for the right reservoir we get
$X_{p'}=h'(\o)+k' G^+_{p'p'}(\o) x_N(\o)$, the
noise statistics of $h'(\o)$ being now determined by $\beta'$. The
left and right reservoirs are independent so that $\la h(\o)
h'(\o') \ra = 0$.  
We can now obtain the particular solution of Eq.~(\ref{seqm}) by
taking Fourier transforms and plugging in the forms of $h(\o)$ and
$h'(\o)$. We then get:
\bea
x_l(t) &=& \int_{-\infty}^{\infty} \zh^{-1}_{lm}(\o) h_m (\o) e^{i \o t}
\label{sol} \\
\zh &=& \hat{\phi}_{lm}-\hat{A}_{lm} ~~~{\rm with} \nn \\
\hat{\phi}_{lm} &=& -(\d_{l,m+1}+\d_{l,m-1})+(2-m_l \o^2) \d_{l,m} \nn \\
\hat{A}_{lm} &=& \d_{l,m} [k^2 G^+_{pp}(\o) \d_{l,1}+k'^2
G^+_{p'p'}(\o) \d_{l,N} ] \nn \\
h_l(\o) &=& k h(\o) \d_{l,1}+ k' h' (\o) \d_{l,N}.  \nn
\eea
We can now proceed to calculate steady state values of observables of
interest such as the heat current and the temperature profile. 
We first need to find the appropriate operators corresponding to
these. To find
the current operator $\hat{j}$ we first define the local energy
density $u_l =\f{p_l^2}{4 m_l}+ \f{p_{l+1}^2}{4 m_{l+1}}
+\f{1}{2}(x_l-x_{l+1})^2$. Using the current conservation
equation $\p \hat{u}/\p t+ \p \hat{j}/\p x =0$ and the equations of
motion we then find that $\hat{j}_l=
( \dot{x}_l x_{l-1} +x_{l-1} \dot{x}_l )/2$.  
The  steady state current can now be
computed by using the explicit solution in Eq.~(\ref{sol}). We get
\bea
\la \hat{j}_l \ra =\int_{-\infty}^{\infty} d \o (i \o) [ k^2 \zh^{-1}_{l,1}(\o)
\zh^{-1}_{l-1,1}(-\o) I(\o)  \nn \\+ k'^2 \zh^{-1}_{l,N}(\o)
\zh^{-1}_{l-1,N}(-\o) I'(\o) ] 
\eea
The matrix $Z$ is tridiagonal and using some of its special properties
(see sec.IA) we can reduce the current expression to the following
simple form:
\bea
\la \hat{j}_l \ra &=& \f{ k^2 k'^2}{\pi}  \int_{-\infty}^{\infty} d \o
\f{\o b (\o) b' (\o)}{\mid Y_{1,N} \mid ^2 } ( f-f') \nn \\ &=&
\int_{-\infty}^{\infty} d \o J(\o) ( f-f')
\label{jsim}
\eea
where $J(\o)=k^2 k'^2 \o b (\o) b' (\o)/{ \pi \mid Y_{1,N}
  \mid ^2 } $ has the physical interpretation as the total heat 
current in the wire due to all right-moving (or left-moving) scattering
states of the full Hamiltonian (system $+$ reservoirs). Such
scattering states can 
be obtained by evolving initial unperturbed states of the reservoirs
with the full Hamiltonian (see end of sec.IA). As before we
have denoted by $Y_{l,m}$ the determinant of the 
submatrix of $\zh$ beginning with the $l$th row and column and ending
with the $m$th row and column. Similarly let $D_{l,m}$ denote the determinant
of the submatrix formed from $\hat{\Phi}$.

\subsection{Ideal reservoirs and contacts:The Landauer case}
For the special case when the reservoirs are also one dimensional
chains with nearest neighbor spring constants $K_{lm}=1$ and the
coupling constants $k,k'$ are set to unity, we have:
$G^+_{pp}=G^+_{p'p'}=e^{-ik}$ where $\o=2 \sin(k/2)$, $I(\o)=f(\o)
\sin(k)/\pi$ for $ \mid \o \mid  <2$ and $I(\o)=0$ for $\mid \o \mid >
2$. In this case  
Eq.~(\ref{jsim})  
simplifies further and has an interpretation in terms of transmission
coefficients of plane waves across the disordered system. We get:
\bea
J &=& \frac{1}{4 \pi} \int_{-2}^{2} d \o  \o |t_N(\o)|^2
(f-f') ~~~~{\rm{where}} \label{curr} \\
|t_N(\o)|^2 &=& \f{4 \sin^2(k)}{|D_{1,N_s}-e^{ik}(D_{2,N_s}+D_{1,N_s-1})+e^{i 2
k}D_{2,N_s-1}|^2}   \nn 
\eea
is the transmission coefficient at frequency $\o$. We have thus
obtained the Landauer formula \cite{land} for phononic transport. It is only in
this special case of a one-dimensional reservoir and perfect contacts
that we get the 
Landauer formula. The reason is that, only in this case is the  
transmission through the contacts perfect, and this requirement is one of
the crucial assumptions in the Landauer derivation. 
Note that in Eq.~(\ref{curr}) (i)the transmission coefficient {\it does
  not} depend on bath properties and (ii) transmission is only through
propagating modes. For general reservoirs where we need to use
Eq.~(\ref{jsim}) the factor $J(\o)$ involves not just the properties
of the wire but also the details of the spectral functions of the
reservoirs. Thus the conductivity of a sample can show rather
remarkable dependence on reservoir properties as we shall see below.  
The above Landauer-like-formula has earlier been stated in \cite{rego}
and derived 
more systematically in \cite{blen}. 
We note that in the high temperature limit $T~,T' \to \infty$
Eq.~(\ref{curr}) reduces to the classical limit obtained exactly in
\cite{dhar,conn,rubin}.  

\subsection{Asymptotic system-size dependences} 
In the case of electrical
conduction the conductance of a long disordered chain decays
exponentially with system size as a result of localization of
states. In the case of 
phonons the long wavelength modes are not localized and can
carry current. This leads to power-law dependences of the current on
system size as has been found earlier in the context of heat
conduction in classical oscillator chains. A surprising result is that
the conductivity of such disordered chains depend not just on the
properties of the chain itself but also on those of the reservoirs to
which it is connected. It can be shown \cite{dhar} that the asymptotic
properties 
of the integral in Eq.~(\ref{jsim}) depend on the low frequency ($\o
\lesssim 1/N^{1/2}$) properties of the integrand. This means that we
will get the same behaviour as in the classical case. We summarize
some of the main results:

(i) The classical case where the    reservoirs are themselves
 one-dimensional. 
In  this case we put $k=k'=1$ and the spectral function
 $G^+_{pp}=G^+_{p'p'}=e^{-ik}$ where $\o=2 \sin (k/2)$.
This was treated by Rubin and Greer \cite{rubin} and it
 was found that    that $J \sim \f{1}{N^{1/2}}$. Thus the ideal
 Landauer case will also show this behaviour. 

(ii) The case of reservoirs which give delta-correlated Langevin noise
corresponds to taking $k=k'=1$ and $G^+_{pp}=G^+_{p'p'}=-i \gamma \o$. The
classical case was first treated by Casher and Lebowitz
\cite{cash,conn} and one gets $J \sim \f{1}{N^{3/2}}$.

(iii) In general one gets $J \sim \f{1}{N^{\alpha}}$ where $\alpha$
depends on the low frequency behaviour of the spectral functions
$G_{pp}(\o)$ and $G_{p'p'}(\o')$ \cite{dhar}.  

Note that the case $\alpha < 1$ leads to infinite thermal conductivity
while $\alpha > 1$ gives a vanishing conductivity.  Thus, depending on
the properties of the heat baths, the same wire can show
either superconducting or   insulating behaviour. 
The usual Fourier's law would predict $J \sim 1/N$, independent of
reservoirs. Thus Fourier's law is not valid in quantum harmonic chains,
even in the presence of disorder. This breakdown of Fourier's law in
$1D$ systems has been noted in a number of earlier studies on
classical systems \cite{four} which have looked at the effects of
scattering both due to  impurities and nonlinearities. 

\subsection{Temperature profiles}
\begin{figure}
\includegraphics[width=\figurewidth]{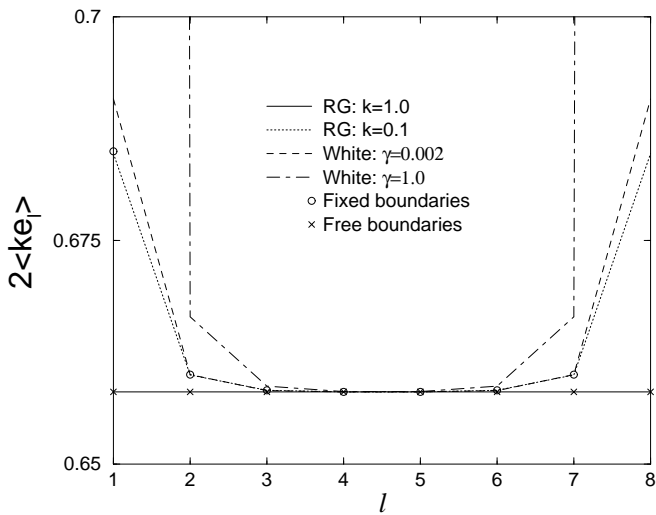}
\caption{ Kinetic energy density profile in a pure harmonic chain
  ($N=8$) attached to reservoirs  at equal temperatures $T=T'=0.2$. Two
different kinds of reservoirs are considered: one-dimenional
  reservoirs (RG) and delta-correlated noise reservoirs (white). The
  exact equilibrium density profiles for an infinite chain (free) and
  one with fixed ends are also given. 
\label{eqpr}}
\end{figure}

 The local temperature of a particle can be
determined from  
its average kinetic energy , $ke_l = \la p_l^2 /(2m_l) \ra$. We get
\bea
ke_l=\f{1}{2} \int_{-\infty}^{\infty} d \o m_l \o^2 [k^2 \zh^{-1}_{l,1}(\o)
\zh^{-1}_{l,1}(-\o) I(\o) \nn \\ +  k'^2 \zh^{-1}_{l,N}(\o)
\zh^{-1}_{l,N}(-\o) I'(\o) ] 
\label{kepr}
\eea
This is straightforward to evaluate numerically for given systems and
reservoirs. For the special case of heat
transmission through a perfect one-dimensional harmonic 
chain attached to one-dimensional reservoirs through  
perfect contacts ({\it i.e} $k=k'=1$) Eq.~(\ref{kepr}) simplifies (for
large $N$) to   
\bea
ke_l=\f{1}{8 \pi} \int_{-\pi}^{\pi} dk \o_k [f(\o_k)+f'(\o_k)]
\label{kepri}. 
\eea
where $\o_k=2 \sin (k/2)$. For $T=T'$, we get:
\bea
ke_l=\f{1}{4 \pi} \int_0^\pi dk \o_k \coth (\f{\beta \o_k}{2})   
\eea
which is the expected equilibrium kinetic energy density on an
infinite chain. For weak coupling to the reservoirs, which can be
achieved by making $k$ and $k'$ small, we expect that the energy
density profile for the system should correspond to that of a finite
chain. We verify this numerically by evaluating Eq.~(\ref{kepr}) for 
$k=k'=0.1$ and $T=T'$ [Fig.~(\ref{eqpr})]. We compare this with the
equilibrium kinetic energy profile of a finite chain given by:
\bea
ke_l=\f{1}{4}\sum_s \o_s \coth (\f{\beta \o_s}{2}) \psi^2_s(l)
\eea
where $\psi_s(l)=[2/(N+1)]^{1/2} sin(k l)$, $\o_s=2 
\sin(k/2)$ where $k=s \pi/(N+1)$ with $s=1,2,...N$.
Note that unlike in the classical case where the energy density is a
constant, in the quantum case, this is not always true.  
It is instructive to look at the equilibrium properties for the case
where the driving is by a delta-correlated noise ( case (ii) discussed
earlier ). In this case the weak coupling limit corresponds to taking
the damping constant $\gamma << 1$. The temperature profiles obtained
from Eq.~(\ref{kepr}) for two different values of $\gamma$ are
plotted in Fig.~(\ref{eqpr}). 

\begin{figure}
\includegraphics[width=\figurewidth]{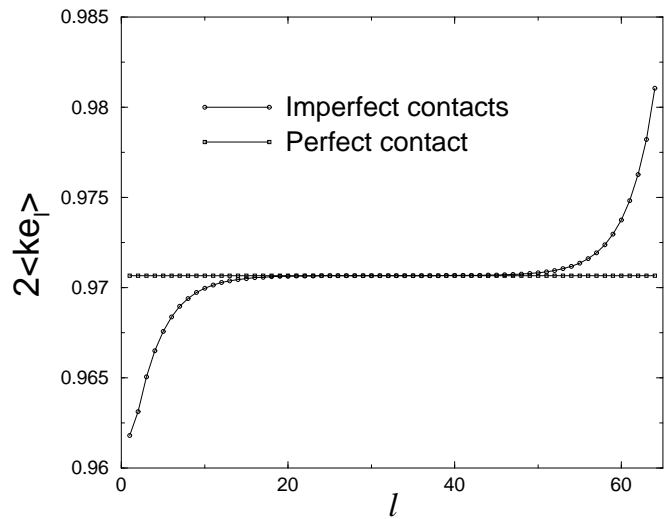}
\caption{ Kinetic energy density profile in a pure harmonic chain
  ($N=64$) attached to one dimensional reservoirs at temperatures
  $T=1.0$ (left) 
and  $T'=0.5$ (right), for perfect and imperfect ($k=k'=0.9$)
  contacts. The temperatures considered are not very high 
and so the bulk temperature is different from the classically expected
value $T_{av}=0.75$. 
\label{tprof}}
\end{figure}

We now consider temperature profiles in the nonequilibrium case ($T
\neq T'$). For the Rubin-Greer ( or Landauer case i.e. $1D$ reservoirs,
perfect contacts ),    
at high temperatures the local temperature is given by $T_l=2 ke_l$
and from Eq.~(\ref{kepri}) we get $T_l=(T+T')/2$ which is the classical
result \cite{ried}. At low temperatures and imperfect contacts $k,k'
\neq 1$ we evaluate the 
local kinetic energy profile numerically using Eq.~(\ref{kepr}). As can be seen
in Fig.~\ref{tprof} the temperature in the bulk still has the same constant
value. At the boundaries however we see a curious feature noted
earlier by \cite{ried,zurcher}: the temperature close to the hot end is {\it
lower} than the average temperature while that at the colder end is
{\it higher} than the average. 
For the case with delta correlated noise, at high temperatures, we
recover the temperature profiles obtained ealier for classical chains
in \cite{ried}. At low temperatures we get results similar to those
found by Zurcher and Talkner \cite{zurcher} and there seem to be some
qualitative differences from the classical temperature profiles,
depending on the value of $\gamma$.  

\section{Discussion} 
 We note that the more popular approach of treating
open quantum 
systems is the Caldeira-Leggett formulation. In that approach,
one deals with density matrices and the treatment becomes
complicated. In the context of the present problem one is not
really interested in the full distribution but rather  
in physical observables like the steady state currents and densities
and these are basically second moments of 
the distribution. The FKM formulation is then more appropriate and
for linear systems one can get exact results. 
The other approach of treating nonequilibrium systems which has been
used quite extensively in the mesoscopic context is
the Keldysh formalism. This is a perturbative treatment where one
writes equations of motion for a set of Green functions and relates them to
self energies through the Dyson equations.
The current is expressed in terms of these Green functions.   In
special cases the 
Dyson equations can be solved exactly and indeed some of our results
can be obtained \cite{caroli,meir,hersh}. On the other hand our method is
more transparent and direct. We integrate out the reservoir degrees of
freedom to get effective Langevin type equations of motion for the
system. These are solved and quickly lead to useful results on
currents and densities of both particle and heat which are
automatically expressible in terms of unperturbed Green functions. 
The connection to  scattering theory is also
immediate and explicit. 
Finally  one obtains a  nice physical picture of the reservoirs serving as
effective sources of noise and dissipation. 
Note that our approach makes 
connections between different approaches such as Caldeira-Legett, Keldysh, 
scattering theory and the transfer-Hamiltonian
method. 

The FKM was earlier used in studying heat transport in
classical disordered 
harmonic chains and it is particularly nice that the method can be
extended to the quantum mechanical regime. Earlier results on
classical chains are then obtained as limiting cases. The more general
quantum mechanical results can be expressed in forms where one can see 
connections with other approaches such as Landauer, Keldysh, etc.  

The dependence of transport properties of a system on the reservoir
properties is at first glance a surprsing fact and we briefly comment
on this.  
From our usual experience in the macroscopic world, one usually thinks
of the conductivity of a system 
as an intrinsic property, not  dependent on the properties of
reservoirs. Imagine making a measurement of the thermal conductivity
of a wire by putting its ends in contact with heat baths at two
different temperatures and measuring the resulting current. The normal
expectation is that the answer should not depend on the material
properties of the heat baths. And indeed this expectation holds out
true quite often. One physical way of understanding this 
is that, as long as the system (the wire) is a strongly  interacting
system, with good ergodicity properties, then one can expect that, soon
after contact is made with the reservoirs, the ends of the wire would
reach a state of local thermal equilibrium with the reservoirs. This
local equilibrium would be completely determined by just the
temperature of the reservoir and this then drives the current in the
wire. In the mesoscopic domain however there are situations when the
interactions between the  carriers are not strong enough to let
the system reach local equilibrium. And then one finds that the
conducting properties of a wire is no longer intrinsic to the wire but
depends on details of the reservoirs. Thus any calculation of
transport properties would require a detailed modelling of the reservoirs.
An explicit demonstration, of the conditions under which
reservoir-dependence goes away, does not seem to exist at present.

As has been shown here the FKM method works as easily for both
electronic transport in disordered  fermionic wires and thermal
transport in disordered harmonic chains. In both cases we are able to
obtain exact formal expressions for particle and thermal currents and
these have very similar forms. Both depend on details of the reservoir
spectral functions. The usual Landauer case where one 
writes the current in terms of transmission factor of one-dimensional
plane waves is shown to follow, exactly, for the choice of one-dimensional
reservoirs and perfect contacts. 
In general however one needs to use modified Landauer formulas and
this can be  quite crucial in interpreting experimental data. For
example we have shown that the 
oscillations in conductance seen in the experiments by Kong et al
cannot be explained unless the  contacts and reservoirs are treated quantum
mechanically.  We also find the rather counterintuitive prediction that
imperfect contacts can enhance the conductance of a wire. 
In the phonon case we make a couple of predictions that are
interesting from the  
experimental point of view: (i) the large system size behaviour of the heat
current  is a power law and the power depends on reservoir properties
(ii) temperature profiles in perfect wires show somewhat
counterintuitive features close to contacts. 
It would be interesting to see if our predictions, which are true for
strictly one-dimensional chains, can be verified in experiments on
nanowires.

We are grateful to N. Kumar, H. Mathur,  T. V. Ramakrishnan and
A. K. Raychaudhuri  for useful comments. A.D acknowledges support from
the National Science Foundation under Grant No. DMR 0086287.
BSS was supported in part by an Indo French grant IFCPAR/2404.1.

{\small $^\dagger$ On leave from Raman Research Institute, Bangalore,
  India.}
   
{\small $^{\star}$ Also at JNCASR, Bangalore, India}

\newpage

\end{document}